\documentclass[twocolumn,aps,superscriptaddress]{revtex4-1}
\usepackage[latin1]{inputenc}
\usepackage{epsf}
\usepackage{amsmath,amssymb}
\usepackage{latexsym}
\usepackage{calc}
\usepackage{color}
\usepackage{graphicx}

\usepackage[normalem]{ulem}

\begin{document}
\title{Embedding via the Exact Factorization Approach}
\author{Lionel Lacombe}
\affiliation{Department of Physics, Rutgers University, Newark, New Jersey 07102, USA}
\author{Neepa T. Maitra}
\affiliation{Department of Physics, Rutgers University, Newark, New Jersey 07102, USA}
\date{\today}
\begin{abstract}
We present a quantum electronic embedding method derived from the exact factorization approach to calculate static properties of a many-electron system. The method is exact in principle but the practical power lies in utilizing input from a low-level calculation on the entire system in a high-level method computed on a small fragment, as in other embedding methods. Here, the exact factorization approach defines an embedding Hamiltonian on the fragment. Various Hubbard models demonstrate that remarkably accurate ground-state energies are obtained over the full range of weak to strongly correlated systems.
\end{abstract}

\maketitle
The computational challenge of performing a quantum calculation of a complex many-body system remains a primary research area in condensed matter physics and quantum chemistry. Density functional theory (DFT) is often turned to, due to its relatively favorable system-size scaling, however limitations of available functional approximations deem it inaccurate for strongly correlated systems. When even DFT gets too expensive for a system of more than a thousand atoms or so, a collection of DFT calculations on subsystems with  functionals modified by couplings to the rest of the system can be used~\cite{Cortona91,WW93,KSGP15,LWMM19}, however such an approach remains inadequate for problems involving strong correlation. 
Instead, one approach in recent years has been to use some kind of quantum embedding method  where the full system is described as an ensemble of two or more fragments:
On the one hand, when the fragments are chosen to be weakly-interacting with each other, the essential idea is that properties of the total system can be obtained by a high level calculation on one modified by input from the other.
 This can be particularly useful when only part of the system is actually of interest, or is strongly-correlated, but such that its environment affects its behavior, and the idea is to calculate accurately properties of the system of interest without having to compute the full problem accurately. 
 On the other hand, when the entire system is of interest and requires a better description of correlation than provided by density functional approximations, GW, and the like, then the high-level calculation can be done successively on different fragments in a self-consistent way, to get a full description of the entire system from several smaller calculations. Several different approaches have been developed in recent decades;
 ranging from the basic embedding variable being the Green's function~\cite{GK92,GKKR96} or directly the self-energy~\cite{KGZ15}, to the density-matrix~\cite{KC12,KC13,SC16} or density~\cite{BSD14}, as well as density-functional based embeddings~\cite{F15}.

Here, we develop a novel embedding method based on the exact factorization (EF) approach. 
EF separates the wave function into a single correlated product of a marginal and a conditional wave function~\cite{Hunter_IJQC1974, Hunter_IJQC1975_2,H86,AMG10,GG14,AMG12}.
Most of the previous EF work focused on separating the electronic from the nuclear part of a molecular wave function, providing an ``exactification" of the Born-Oppenheimer (BO) approximation:
$
  \Psi(\bold{R}_1,\bold{R}_2..;\bold{r}_1,\bold{r}_2..) = 
  \chi(\bold{R}_1,..)\Phi_{\bold{R}_1,..}(\bold{r}_1,..)
$
where the marginal, $\chi(\bold{R}_1,..)$, is the nuclear wave function and $\Phi_{\bold{R}_1,..}(\bold{r}_1,..)$ the electronic part parametrized by nuclear coordinates.
This approach has been successful for giving insight into effects of electronic-nuclear coupling on dynamics (e.g. Refs.~\cite{AMG12,AASG13})  as well as in  deriving practical non-adiabatic quantum-classical methods~\cite{MAG15,AMAG16,MATG17,HLM18,AC19,FMC19,FPMC19}. There have been generalizations in several directions; most notably for the present purposes are the exact single-active electron approach arising from   factorizing a purely electronic wave function  into a one-electron marginal and the rest~\cite{H86,SG17}, and the formal generalization to arbitrary many-body non-real-space Hamiltonians~\cite{GZR18}. 

The present work extends the EF approach to a completely new class of applications.  
In our embedding via the exact factorization (EVEF) approach, we factorize the full electronic wave function in Fock space. 
The idea is to solve the full system with a low-level calculation (e.g. Hartree-Fock (HF)), and use the solution to generate an approximate Hamiltonian for the marginal corresponding to a fragment which is then solved with a high-level method (e.g. exact diagonalization).
The fragment is a chosen set of single-particle orbitals in the basis defining the Fock space.; for example, these can be selected to be the more strongly-correlated orbitals in the one-electron Hilbert space. We present three levels of EVEF, each increasingly refined, and the sensitivity to the choice of fragment depends on which level is chosen. 
The results on different Hubbard systems show that EVEF is able to capture the range from weak to strong correlation in an efficient and accurate way.

The Fock space electronic wave function in a space of $M$ single-particle orbitals, $\Psi(n_1...n_M)$, is defined via $\vert \Psi \rangle = \sum_{n_i=\{0,1\}} \Psi(n_1...n_M)\vert n_1...n_M\rangle$, where $\vert n_1...n_M\rangle$ represents a single Slater determinant with $n_i = 0$ or $1$ representing the occupation of spin-orbital $i$. 
Choosing the  first $K$ spin-orbitals to span the fragment space, the factorization reads:
\begin{equation}
  \Psi(\underbrace{n_1,n_2,..n_K}_{\underline{n}}
  ,\underbrace{n_{K+1},..n_{M}}_{\underline{m}}) = 
  \chi(\underline{n})\Phi_{\underline{n}}(\underline{m})
  \label{eq:EF_fock}
\end{equation}
where the marginal wave function
 $\chi(\underline{n})$ is a function of the fragment configuration and $\Phi_{\underline{n}}(\underline{m})$ is the conditional part. 
 The factorization is unique up to an $\underline{n}$-dependent phase,$F(\underline{n})$, provided the Partial Normalization Condition (PNC),
 \begin{equation}
  \sum_{\underline{m}}
  \Phi^{*}_{\underline{n}}(\underline{m})
  \Phi_{\underline{n}}(\underline{m})
  = 1
  \label{eq:PNC}
\end{equation}
is satisfied, adapting the proof of Ref.~\cite{AMG10,AMG12}. Then, it follows that
$
    \chi(\underline{n}) = \sqrt{\sum_{m_j={0,1}}
    |\Psi(\underline{n},\underline{m})|^2}\times(e^{iF(\underline{n})})
  $
(with $m_j$ going over $K+1$ to $M$ spin-orbitals) where the arbitrary phase factor $e^{iF(\underline{n})}$ represents the so-called gauge freedom of EF.
The EF approach usually proceeds by finding a coupled set of equations for the marginal and conditional factors, which contain terms that exactly account for coupling of the two subsystems. However, at this point we deviate from what is usually done in EF: here, we find an equation for $\chi$ that emulates the effect of $\Phi_{\underline{n}}$ without ever having to solve the numerically challenging non-linear and non-Hermitian equation for the conditional wave function~\cite{GLM19,GZR18}. 

To obtain the equation for $\chi(\underline{n})$, consider first the full Schr\"odinger equation for $\Psi$, which involves the full exact Hamiltonian $\hat{H}$:
\begin{equation}
  \sum_{\underline{n}',\underline{m}'} 
  H_{\underline{n},\underline{m};\underline{n}',\underline{m}'}
  \Psi_{\underline{n}',\underline{m}'}
  =E\Psi_{\underline{n},\underline{m}}
\end{equation}
where $H_{\underline{n},\underline{m};\underline{n}',\underline{m}'} = \langle \underline{n},\underline{m} \vert \hat{H} \vert \underline{n}',\underline{m}' \rangle$. 
Inserting the factorized form Eq.~(\ref{eq:EF_fock}), multiplying on the left by $\Phi^{*}_{\underline{n}}(\underline{m})$, and summing over $\underline{m}$ gives our eigenproblem for $\chi(\underline{n})$:
\begin{equation}
  \sum_{\underline{n}'}
  h_{\underline{n};\underline{n}'}\chi(\underline{n}')
  =
  E\chi(\underline{n})
  \label{eq:embedded_eigen}
\end{equation}
where we have used the PNC Eq.~(\ref{eq:PNC}) on the right-hand-side, and identified the embedded Hamiltonian
\begin{equation}
  h_{\underline{n};\underline{n}'}
  \equiv
  \sum_{\underline{m}',\underline{m}}\Phi^{*}_{\underline{n}}(\underline{m})
  H_{\underline{n},\underline{m};\underline{n}',\underline{m}'}
  \Phi_{\underline{n}'}(\underline{m}')
  \label{eq:emb_H}
\end{equation}
So far, everything is exact, and $E$ could be any  eigenvalue of the full Hamiltonian; it need not be the ground-state energy.
If it was possible to somehow obtain this embedded Hamiltonian exactly, then the exact ground-state energy of the full system could be obtained by solving the eigenproblem 
of Eq.~(\ref{eq:embedded_eigen}) in the small Hilbert space of just the fragment, regardless of how small it is, even for a single-orbital fragment! 
Further, if we could somehow obtain the embedded observable, $o_{\underline{n};\underline{n}'} = \sum_{\underline{m'},{\underline{m}}} \Phi^*_{\underline{n}}(\underline{m})
  O_{\underline{n},\underline{m};\underline{n}',\underline{m}'}
  \Phi_{\underline{n}'}(\underline{m}')$, for any many-body operator on the full system $\hat{O}$, then the solution of Eq.~(\ref{eq:embedded_eigen}) yields the expectation value of $\hat{O}$, through $\langle\Psi\vert\hat{O}\vert\Psi\rangle = \sum_{\underline{n};\underline{n}'} \chi^*({\underline{n}})o_{\underline{n};\underline{n}'}\chi(\underline{n'})$.

Finding the exact embedded Hamiltonian Eq.~(\ref{eq:emb_H})  is of course as hard as solving the original problem. The practical power of this set-up depends on making an approximation, so this enters in
the first step in our EVEF approach.
We solve the HF Hamiltonian $\hat{H}^{MF}$ for the whole system first to obtain the HF state:
\begin{equation}
  |\Psi^{MF}\rangle = \prod_j \left(\sum_i C_{i,j} \hat{a}^{\dagger}_i \right)|~\rangle
  \label{eq:HF_Psi}
\end{equation}
where $\hat{a}^{\dagger}_i$ is the creation operator in a given single-particle basis (e.g. the site-basis in lattice models), and $|~\rangle$ is the vacuum state.
In the second step, the embedded Hamiltonian  Eq.~(\ref{eq:emb_H})  is computed using $\Phi^{MF}_{\underline{n}}(\underline{m})=\Psi^{MF}(\underline{n},\underline{m})/\chi^{MF}(\underline{n})$. For configurations ${\underline{k}}$ where $\chi^{MF}(\underline{k}) = 0$,  $\Phi^{MF}_{\underline{k}}(\underline{m})$ becomes ill-defined and  is set to zero. 
Using the resulting mean-field-derived embedded Hamiltonian to solve Eq.~(\ref{eq:embedded_eigen}) exactly (or with a high-level method) gives us directly an approximation for the total energy. In the Supplemental Material, we rewrite the entire EVEF formalism in second quantization which allows us to compute the embedded Hamiltonian using Wick's theorem.

The two steps above describe the central approach of this paper and it is what we call EVEF-1. It  can be recast as the minimization problem:
\begin{equation}
  E \approx
  \min_{\chi,\vert\vert\chi\vert\vert = 1}  
  \sum_{\underline{n}',\underline{m}',\underline{n},\underline{m}}
  \chi^{*}(\underline{n})\Phi^{MF\,*}_{\underline{n}}(\underline{m})
  H_{\underline{n},\underline{m};\underline{n}',\underline{m}'}
  \Phi^{MF}_{\underline{n}'}(\underline{m}')\chi(\underline{n}')
\end{equation}
with $\Phi^{MF}$ kept fixed. After this minimization the resulting wave function $\chi(\underline{n})\Phi^{MF}_{\underline{n}}(\underline{m})$ is typically not a single Slater determinant, that is, the procedure introduces some correlation, and the resulting energy lies between the mean-field result and the exact energy.

EVEF-1 is most effective in giving a significant energy correction when, for most fragment configurations, $\chi^{MF}({\underline{n}})$ are non-zero. This can be understood from realizing that configurations ${\underline{k}}$ for which $\chi^{MF}(\underline{k}) = 0$ do not contribute to the computation of the energy, since the zeroing of the corresponding conditional part results in $h_{\underline{k},\underline{n}'} = h_{\underline{n},\underline{k}} = 0$. 
An extreme case would be if one performs EF in the basis of HF orbitals. In this case $\chi^{MF}(\underline{n})$ is zero for all $\underline{n}$ except for the one corresponding to the configuration of occupied HF orbitals and the matrix $h_{\underline{n},\underline{n}'}$ reduces to a $1\times1$ matrix equal to the HF energy.
In contrast, EVEF-1 is very effective in a case where $\chi^{MF}(\underline{n})$ has a similar amplitude for every $\underline{n}$, and $\underline{n}$ are chosen such that there is strong correlation between the orbitals of the fragment while the environment reacts to the fragment in a mean-field manner.

In rare cases, however,  the choice of fragment leads to $\chi^{MF}(\underline{n}) = 0$  for a large fraction of configurations. It could even be possible that if
the full system is treated as the fragment, the approximation will not reproduce the exact energy, because of the zeroing of the conditional wavefunction for configurations where $\Psi^{MF}(\underline{n}) = 0$. This implies that $\vert \Psi^{MF} \rangle$ is a bad starting point and a higher-level approach for the initial guess of the full wave function is needed.  A pedagogical illustration of the choice of the fragment, a half-filled two-site Hubbard model,  is given in the Supplemental Material. We expect this kind of situation to be less likely in the case of a bath large compared to the fragment size.   

EVEF-1 is expected to work well when only one subsystem (chosen as the fragment) is strongly-correlated while the rest is well-described by a mean-field. In the case of strong correlation throughout, we instead partition the system as an ensemble of say $N_f$ non-overlapping fragments, each of which is computed with a high-level method using an embedded Hamiltonian generated from HF,  and combine the results.
We describe this method, EVEF-2, next. 

In an exact calculation, each and any fragment would correspond to a different embedded Hamiltonian  in Eq.~(\ref{eq:embedded_eigen}), denoted now $\hat{h}_\alpha$ where $\alpha$ labels the fragment, yet each would yield
 the same eigenvalue $E$, the total energy of the system.
 However, when an approximate $\Psi$ is used, each fragment gives a different answer for the  energy. 
Moreover, in any fragment calculation, some contributions to the matrix elements of $\hat{h}_\alpha$  yield no new correlations beyond HF. To see this, consider a Hamiltonian of the form
\begin{equation}
  \hat{H} =
  \sum_{i,j} t_{ij} \hat{a}^{\dagger}_i \hat{a}_j
  + \sum_{i,j,k,l} w_{ijkl} \hat{a}^{\dagger}_i \hat{a}^{\dagger}_j \hat{a}_k \hat{a}_l
\end{equation}
where $w_{ijkl} = \langle i\,j\vert \hat{W}\vert l\,k \rangle$.
If none of $i,j,k,l$ are contained in fragment $\alpha$, that term contributes only to the diagonal of $\hat{h}_\alpha$ and with no corrections to its HF value, but it does yield non-trivial correlations in other fragments. This suggests that a better approach would be to extract, for each fragment, only the part of the energy altered by the high-level method, and then sum this over fragments. 
That is, we partition the full Hamiltonian as a sum of fragment contributions
$
  \hat{H} = 
  \sum_{\alpha} \hat{H}^{\rm loc}_{\alpha}
$
where $\hat{H}^{\rm loc}_{\alpha}$ is ``local", defined by 
\begin{equation}
  \hat{H}^{\rm loc}_{\alpha} = 
  \sum_{i \in \alpha,j} t_{ij} \hat{a}^{\dagger}_i \hat{a}_j
  + \sum_{i \in \alpha,j,k,l} w_{ijkl} \hat{a}^{\dagger}_i \hat{a}^{\dagger}_j \hat{a}_k \hat{a}_l 
\end{equation}
where only terms with the first index inside the fragment are included.
We then define a ``local" embedded matrix $h^{\rm loc}_\alpha$ for a fragment $\alpha$ in an environment consisting of the other $(N_f - 1)$ fragments:
\begin{equation}
  h^{\rm loc}_{\alpha;\underline{n}_\alpha;\underline{n}'_\alpha}
  =
  \sum_{\underline{m}'_\alpha,\underline{m}_\alpha}\Phi^{*}_{\alpha;\underline{n}_\alpha}(\underline{m}_\alpha)
  H^{\rm loc}_{\alpha;\underline{n}_\alpha,\underline{m}_\alpha;\underline{n}_\alpha',\underline{m}_\alpha'}
  \Phi_{\alpha;\underline{n}_\alpha'}(\underline{m}_\alpha')
  \label{eq:emb_Hloc}
\end{equation}
(which is similar in spirit to the Density-Matrix Embedding Theory (DMET) fragment energy defined in Ref.~\cite{KC13}).
The total energy is obtained by first computing the energy of each fragment $\alpha$ using $\chi_\alpha$, where $\chi_\alpha$ is the solution of Eq.~(\ref{eq:embedded_eigen}) with $h_\alpha$ on the left, and then summing over all the fragments that form the full partition of the system:
\begin{equation} 
  E = \sum_{\alpha} E_\alpha
  \quad\text{with}\quad
  E_\alpha = \chi^{\dagger}_{\alpha} h^{\rm loc}_\alpha\chi^{\phantom{\dagger}}_{\alpha}
  \label{eq:partition_total_energy}
\end{equation}
In the case no approximation is made for $h_\alpha$ or $h^{\rm loc}_\alpha$, this energy
is exact and equal to the $E$ appearing in Eq.~(\ref{eq:embedded_eigen}). 
One can partition any many-body observable $\hat{O}$ in the same way as a sum of $\hat{O}_\alpha^{\rm loc}$, embedding it  analogously to Eq.~(\ref{eq:emb_Hloc}), i.e. 
$
\langle \Psi \vert \hat{O} \vert \Psi \rangle = \sum_\alpha \chi^{\dagger}_{\alpha} o^{\rm loc}_\alpha\chi^{\phantom{\dagger}}_{\alpha}
$.

Of course, in practice, an approximation is used, and the steps then for EVEF-2 are as follows: First, as in EVEF-1, the HF problem for the whole system is solved yielding Eq.~(\ref{eq:HF_Psi}), and for each fragment $\chi_\alpha^{HF}$ is computed in terms of $C_{i,j}$ in the same way. In the second step, for each fragment, the embedded Hamiltonian $h_\alpha$ is computed from Eq.~(\ref{eq:emb_H}) using  $\Phi^{MF}_\alpha$  and the full $\hat{H}$, and Eq.~(\ref{eq:embedded_eigen}) is solved for each fragment $\alpha$: $h_\alpha\chi_\alpha = \tilde{E}\chi_\alpha$ to find $\chi_\alpha$. Third, the local matrix $h^{\rm loc}_\alpha$ is formed using $|\Psi ^{MF}\rangle$ in Eq.~(\ref{eq:emb_Hloc}), and the fragment energy $E_\alpha$ and total energy are computed from Eq.~(\ref{eq:partition_total_energy}). 
Unlike EVEF-1 however, EVEF-2 does not provide a wave function for the whole system and is not variational, so, like in DMET and Dynamical Mean Field Theory (DMFT), the EVEF-2 energy may fall below the exact ground-state energy. 

A refinement of EVEF-2 follows from introducing a self-consistency criterium and modifying
the mean field with a local or a non-local potential to fit an observable in each fragment. This is in a similar spirit to what is done in DMET and DMFT. Here, in EVEF-3, we consider fitting the orbital occupation. The procedure follows that of EVEF-2, but at the end of the second step we include a chemical potential on the fragment orbitals $\mu_j$, $j \in \alpha$ to minimize $||n_j-n^{MF}_j|| ^2$ where $n_j$ is the average occupation of orbital $j$ that can be directly obtained from $\chi(\underline{n})$ using $n_j=\sum_{\substack{n_i = \{0,1\}\\n_i \neq n_j}}        |\chi(\underline{n})|^2$. We then iterate the first two steps until convergence is obtained.


To test our approach, we computed the energy in different Hubbard systems, with a general Hamiltonian 
\begin{equation}
  \hat{H} = -\sum_{<ij>,\sigma} t_{ij}\hat{a}^{\dagger}_{i,\sigma}\hat{a}_{j,\sigma} + \sum_{i} U_{i} \hat{n}_{i,\uparrow}\hat{n}_{i,\downarrow}
\end{equation}
There is no local potential but the on-site repulsion may vary from site to site.
We set the gauge freedom $F(\underline{n})$ to zero, but a few tests with a different choice showed that it made no difference here.
All units are arbitrary.

\begin{figure}
\includegraphics[width=1.0\columnwidth]{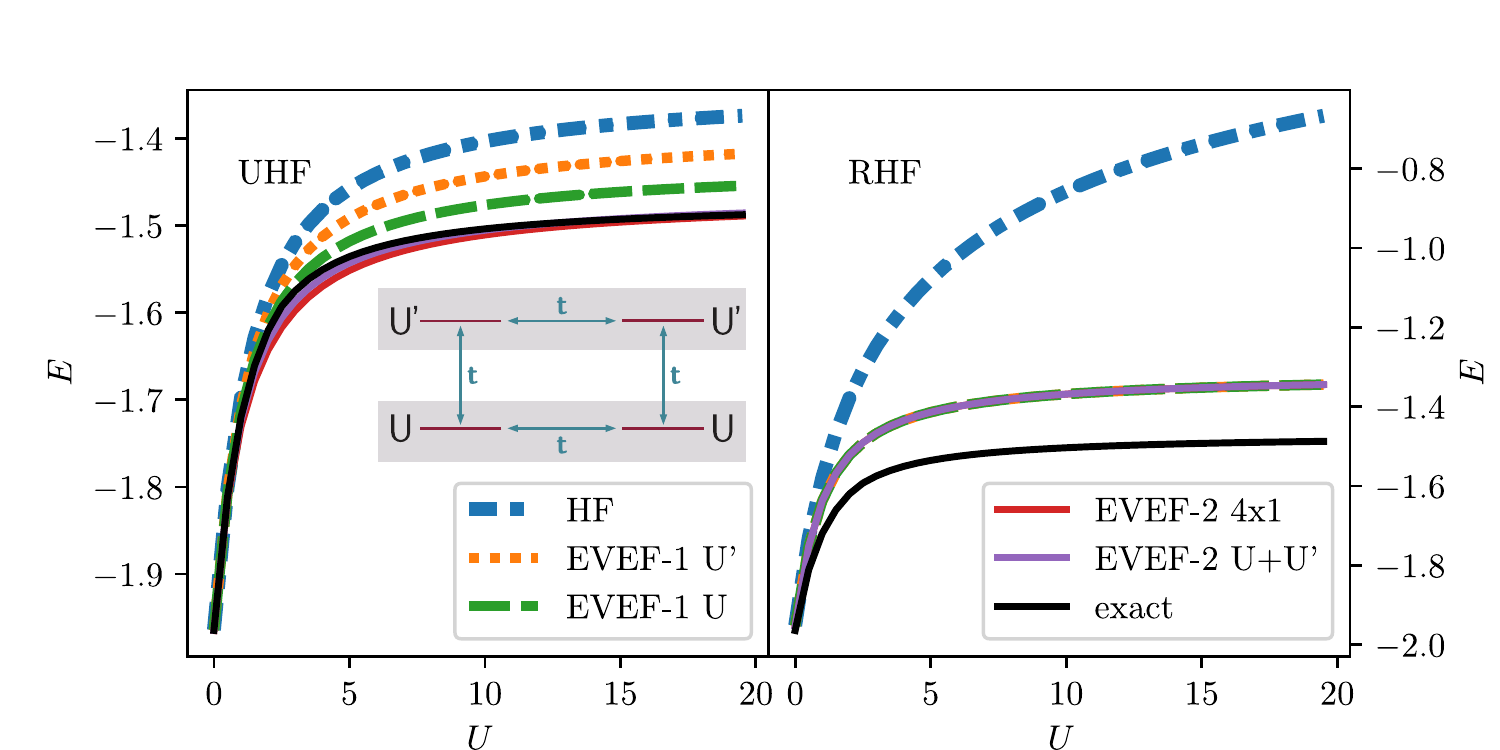}
\caption{Hubbard tetramer, sketched in the inset.  We take $t = 1/2$, $U' = 0.1$ and $U$ as variable.
  Total energy $E$ as a function of $U$, for a UHF full-system calculation (left) and RHF (right), using EVEF-1 or EVEF-2 with fragments as indicated in the legend.
}
\label{fig:e_tetramer}
\end{figure}

Our first system is a molecule represented by the {\it Hubbard tetramer} depicted in the inset in Fig.~\ref{fig:e_tetramer}, with a variable on-site repulsion $U$ on two of the sites, while the other two-sites are weakly-interacting  with a fixed $U' = 0.1$. 
Results of EVEF-1 and EVEF-2 for the total energy $E$ as a function of $U$ are shown in Fig.~\ref{fig:e_tetramer}. 

Consider first the left panel that corresponds to the calculation using Unrestricted HF (UHF) for $\Psi^{MF}$. 
The exact curve (black solid line) is indistinguishable from HF at small $U$ since the total correlation from $U$ and $U'$ is small, and it saturates quickly for $U\approx 5$, becoming nearly constant with $E=-1.5$. 
UHF (blue dash-dotted) follows the same trend but saturates a little later and at higher energy, around $E=-1.38$. 

The green dashed curve is the result from EVEF-1 with the natural choice of fragment being the two sites with local repulsion $U$. This gives a dramatic improvement over UHF for intermediate and strong correlations $U$. 
One can also make the counterintuitive choice to treat the $U'$ part at higher level in the bath of the $U$ sites (orange dotted line). 
Interestingly, this also considerably improves  the energy at large $U$ even though the correlation is almost entirely in the two sites that are {\it not} treated at the higher level. 
The effect of $U$ is partially contained in the definition of the embedded Hamiltonian $h$ and the ensuing diagonalization brings back some of the correlation.

The two previous fragment calculations can also be used as partition for EVEF-2 (grey rectangles in the inset of Fig.~\ref{fig:e_tetramer})).
Doing so gives a remarkably accurate energy (the violet solid curve), which is almost on top of the exact result. 
Another possibility is to treat each site as an independent fragment as a $4 \times 1$ partition for EVEF-2.  
This is represented by the solid red curve, very close to the violet and black ones (but slightly worse around $U = 5$).

The right panel of Fig.~\ref{fig:e_tetramer} instead takes $\Psi^{MF}$ as Restricted HF (RHF).
In this case all EVEF (1,2,3) approaches are equivalent and, although a significant improvement over RHF is obtained especially at large $U$, they generate the same energy, lying  between the exact and RHF results.
In the case of RHF, the HF determinant provides a $\chi^{HF}(\underline{n})$ which is zero for too many configurations $\underline{n}$ (see earlier discussion) that are significant for the exact $\chi$. 
The Hilbert space in which $h$ is actually diagonalized is too small and is the limiting factor for improvement of the energy.

\begin{figure}
\includegraphics[width=1.0\columnwidth]{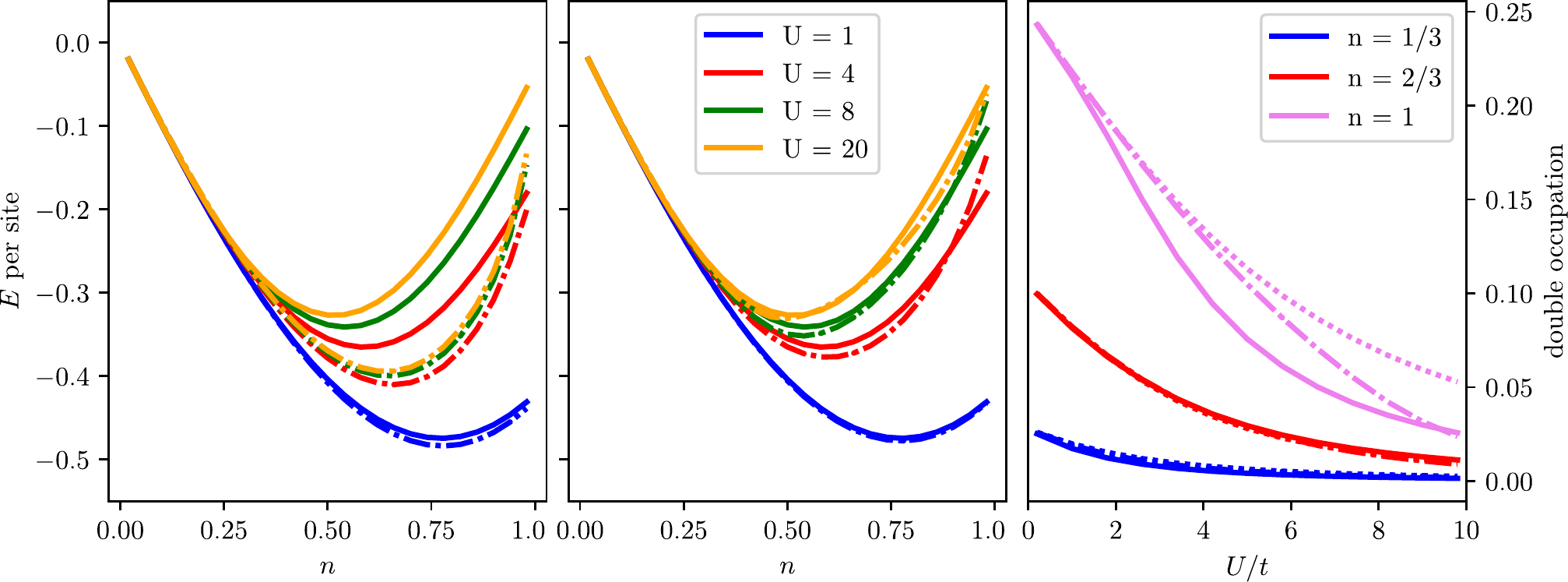}
\caption{
  Energy per site as a function of the occupation per site $n$ in a Hubbard ring of $100$ sites for $U$-values indicated. Left panel shows exact (solid) and EVEF-3 using $1$-site fragment (dash-dotted), middle panel uses the $2$-sites fragment (dash-dotted). The right panel shows double occupation $\langle \hat{n}_{i\uparrow} \hat{n}_{i\downarrow}\rangle$ as a function of the ratio $U/t$ for the three different $n$ indicated; exact (solid), EVEF-3 with  $1$-site fragment (dotted) and $2$-sites fragment (dash-dotted). }
\label{fig:e_chain_100}
\end{figure}

Our next model system is the {\it uniform 100-site Hubbard ring} with $t = 1/2$ and the same $U$ on each site. We calculate the energy from EVEF as a function of filling fraction $n$ per site for different values of $U  = 1,4, 8, 20$ and compare with the Bethe-Ansatz solution for an infinite chain~\cite{S72}  as an approximately exact reference. 

The results shown in Fig.~\ref{fig:e_chain_100} are from EVEF-3 using RHF as the mean-field; we found that without a chemical potential (as in EVEF-2), the number of electrons in the fragment turned out unphysical; for example, as $n\to 0$, the average number of electrons in the fragment did not go to zero. 
The left panel takes the fragments to have one site while the middle panel has two-site fragments. Because of the homogeneity of the system, only one fragment calculation (one $h^{\rm loc}$) needs to be done and only one chemical potential $\mu$ is needed.

In both cases, EVEF-3 produces very good results for small and intermediate $U$, but is increasingly worse  at larger $U$ for the 1-site fragment. On the other hand, the 2-site fragment calculation greatly improves the energy, making the curve very close to the exact one even for very strong interaction strengths $U$.
The worst results are obtained when approaching $n = 1$ where both the derivative and the value of the energy are overestimated.


As another observable, we display, in the third panel, the double occupation in the site-basis, $\langle \hat{n}_{i\uparrow} \hat{n}_{i\downarrow}\rangle$. This is directly available from $\chi$, e.g. as  $\langle \hat{n}_{i\uparrow} \hat{n}_{i\downarrow}\rangle=\vert \chi(1,1) \vert^2$ when using a $1$-site fragment.  
For intermediate filling $n$, both $1$-site and $2$-sites fragments generate almost the exact result; the $1$-site fragment calculation is even better for this observable than it is for the total energy. 
The EVEF error is greater at half filling $n=1$,  consistent with the larger energy error at this $n$.

In comparison with DMET calculations for this system~(Figs 1 and 2 of Ref.~\cite{KC12}), 
a similar deviation is seen. In fact, while DMET performs better for the energy but worse for double-occupation than $1$-site fragment EVEF, EVEF appears to outperform DMET 
 for  the $2$-site fragment at larger $U$. Since the $n$-site fragment calculation in DMET requires the high-level calculation in a $2n$-site Hilbert space, then one could argue that  the EVEF $2$-site fragment calculation should be compared against the $1$-site DMET one, and in such a comparison the errors in EVEF are much less. 
Calculations for a larger fragment/system size will likely improve the results further and are left for future work.


In summary, we have derived a practical embedding method from the EF approach, establishing a new class of applications for the EF idea. We proposed three levels of refinement, EVEF-1, -2 and -3.
The formalism  is general enough to be applied directly to any quantum system; for example, to study molecular dissociation, metal-insulator transitions, transition metal oxides, stripe/superconducting phases, and through projection, any fragment observable could in theory be obtained.
The method produces results that are quantitatively good when tested on different Hubbard systems: a tetramer and a uniform ring, for the full range from weak to strong correlation.
The accuracy is comparable to other embedding methods like DMET, and in some cases better, but 
the Hilbert space of the fragment in our approach is smaller. 
As in DMET, EVEF is based on a wavefunction rather than the Green's function that DMFT and self-energy embedding theory are based on, and this may have practical advantages due to using a frequency-independent quantity.  Unlike DMET, the single product form of our wave function enables us to bypass the embedding basis which offers possible further numerical advantage, and it can be straightforwardly applied with approximate wave functions beyond Slater-determinants. One advantage of EVEF is its flexibility as it can be used with any method that provides the expectation values needed in the definition of $h$, and further, it can be directly extended to excited states.
 A detailed comparison with other methods and molecular or solid-state systems is left for future work, as are improvements and extensions, 
 such as improving the
 stability of the self-consistency loop in EVEF-3, choosing different observables to match between the HF and fragment calculations, application to excited states, and a real-time extension via a time-dependent variational principle.

Note added: We recently became aware of work by Requist and Gross \cite{RG19} developing a similar exact factorization-based embedding method.

\begin{acknowledgments}
We thank Ryan Requist for helpful discussions. 
Financial support from the U.S. National Science Foundation CHE-1940333   and the Department of Energy  Office
of Basic Energy Sciences, Division of Chemical Sciences, 
Geosciences and Biosciences under Award DE-SC0020044 are gratefully acknowledged. 
\end{acknowledgments}

\bibliography{./ref_evef}

\end{document}